\documentclass[12pt, preprint]{aastex}
\usepackage{emulateapj5}  

\lefthead{Fabbiano et al.}  \righthead{{\it Chandra} ACIS-S
Observations of NGC 4038/4039}

\def\ergcm2s{~erg cm$^{-2}$ s$^{-1}$ } 
\def\ergs{~erg s$^{-1}$}		
\def\nh{~$\rm{N_{H}}$}

\def\etal{et al.~}		

\def\msun{~M$_{\odot}$}

\def\n4038{~NGC4038/39}		
\def\chandra{{\it Chandra }}

\def\x2{$\chi^{2}$}	



\begin{document}

\title{A Variable Ultraluminous Supersoft X-ray Source in ``The Antennae'':
Stellar-Mass Black Hole or White Dwarf?
 \\}

\author{ G. Fabbiano$^1$, A. R. King$^2$, A. Zezas$^1$, T. J. Ponman$^3$,
A. Rots$^1$ and  Fran\c cois Schweizer$^4$}
\affil{$^1$Harvard-Smithsonian Center for Astrophysics, 60 Garden
Street, Cambridge, MA 02138}
\affil{$^2$Theoretical Astrophysics Group, University of Leicester, Leicester 
LE1 7RH, UK}
\affil{$^3$School of Physics \& Astronomy, University of Birmingham, Birmingham 
B15 2TT, UK}
\affil{$^4$Carnegie Observatories, 813 Santa Barbara St., Pasadena,
CA 91101-1292}

\bigskip

\begin{abstract}
The \chandra\ monitoring observations of The Antennae (NGC~4038/39)
have led to the discovery of a variable, luminous, supersoft source
(SSS). This source is only detected at energies below 2~keV and, in
2002 May, reached count rates comparable to those of the nine
ultraluminous X-ray sources (ULXs) detected in these
galaxies. Spectral fits of the SSS data give acceptable results only
for a $\sim$100--90~eV blackbody spectrum with an intrinsic absorption
column of $N_{\rm H} \sim 2-3 \times 10^{21} \rm cm^{-2}$. For a
distance of 19~Mpc, the best-fit observed luminosity increases from
1.7$\times 10^{38}$\ergs\ in 1999 December to 8.0$\times
10^{38}$\ergs\ in 2002 May. The intrinsic, absorption-corrected
best-fit luminosity reaches 1.4$\times 10^{40}$~\ergs\ in 2002
May. The assumption of unbeamed emission would suggest a black hole of
$\ga$100\msun. However, if the emission is blackbody at all
times, as suggested by the steep soft spectrum, the radiating
area would have to vary by a factor of $\sim$10$^3$,
inconsistent with gravitational energy release from within a few
Schwarzschild radii of a black hole.  Viable explanations for the
observed properties of the SSS are provided by anisotropic emission
from either an accreting nuclear-burning white dwarf or an accreting
stellar-mass black hole.

\end{abstract}


\section{Introduction}

At a distance of 19~Mpc ($H_o = 75$), NGC~4038/39 (The Antennae) have
long been studied as the nearest example of a galaxy pair undergoing a
major merger (Toomre \& Toomre 1972).  In the X-ray band of 0.1--10~keV,
the first {\it Chandra} observation of this system in 1999 December 
revealed an extraordinarily rich population of luminous point-like sources
(Fabbiano et al.\ 2001). We are now in the midst of a year-long {\it
Chandra} monitoring program of The Antennae.  The first results of
this program, on the luminosity and spectral variability of nine
ULXs (see Fabbiano 1989 and Makishima et al.\ 2000 for earlier work on ULXs), 
detected with luminosities $L_X > 10^{39} \rm ergs~s^{-1}$, are
reported in Fabbiano et al.\ (2002).

Here we report the discovery of a very luminous, variable, supersoft
source in the Antennae galaxies. While SSSs with blackbody spectra of
$\sim\,$40--100~eV have been detected in several galaxies (e.g., in
Local Group galaxies: Greiner 1996; M81: Swartz et al.\ 2002; M101:
Pence et al.\ 2001), their typical luminosities do not exceed
$10^{39}$\ergs\ and are mostly lower than $10^{38}$\ergs.
These sources are believed to be the result of nuclear burning on the
surface of accreting white dwarfs in binary systems (van den Heuvel
et al.\ 1992; see also Kahabka \& van den Heuvel 1997).  However, the source
of interest here, CXOANT J120151.6$-$185231.9 (= Source 13 of the Zezas et
al.\ 2002a list), may reach luminosities in excess of $10^{40}$\ergs.

Source 13 was first detected in 1999 December with a count rate
$\sim$4 times below that of the ULX range.  It is indicated by a circle 
in fig.~1. Our monitoring of
The Antennae shows that this source reached a count rate comparable to
those of the ULXs in 2002 May, while keeping an unusual, very
soft emission.  In this paper, we present both the light curve of this
source and a spectral study of its emission.

\section{Observations and Analysis}

Table~1  summarizes the log of the four {\it Chandra} ACIS-S3
(Weisskopf et al.\ 2000) observations of The Antennae discussed in the
present paper, and lists the observing times after screening for
background flares.  Details of the data analysis are given in Fabbiano
et al.\ (2003). This analysis includes
astrometric correction of the 2001 December, 2002 April, and 2002 May
observations to the 1999 December coordinates, 
and source detection in four spectral bands following
the prescriptions by Zezas et al.\ (2002a): Full band (0.3--7.0)~keV;
soft (0.3--1.0)~keV; medium (1.0--2.5)~keV; and hard (2.5--7.0)~keV.
The data were corrected for spatial and spectral variations of the
ACIS-S3 response, including the time-variable ACIS-S3 effective
area{\footnote{http://asc.harvard.edu/cal/Acis/Cal\_prods/qeDeg/}
\footnote{http://www.astro.psu.edu/users/chartas/xcontdir/xcont.html}}.
This correction results in a factor of 1.8 increase of the count rate in the
last three observations, when compared with the 1999 December data.
The back-illuminated ACIS-S3 CCD is not affected by the energy response
degradation experienced by the front-illuminated CCDs; therefore the CTI
correction is not relevant here.

Figure~2 shows the light-curve of Source 13 in the 0.3--7.0~keV band.
We observe a $\sim$8-fold increase of the corrected count rate from
1999 December to 2002 May.  While detected at sub-ULX count rates in the
1999 December data (Zezas et al.\ 2002a), in May 2002 the count rate is
in the range of the ULX count rate in The Antennae (corresponding to
$L_X > 10^{39}$\ergs, for a 5~keV bremsstrahlung spectrum), but the
spectrum is much softer than the spectra of the other ULXs. This
source remains undetected in the hard band, and only the color $(S-M/S+M)$
can be derived.  This color is 0.88$\pm$0.04, significantly softer
than the typical ULX colors that all lie in the 0.2 to $-$0.4 range
(Fabbiano et al.\ 2003).

We used XSPEC for the spectral analysis (see Zezas et al.\ 2002a).
Source counts were extracted from circles of 3~pixel (1.5\arcsec) radius,
with background estimated from surrounding annuli. The data from each
observation were fitted with a variety of models, available in the XSPEC library,
including blackbody, 
disk-blackbody (used in the ASCA studies of ULXs, see Makishima et al. 2000),
bremsstrahlung, Raymond-Smith thin-plasma model (RS), and power-law.
A variable absorption column was included in
each model.  Spectral analysis was performed on data at energies
0.3--7.0~keV for all the data sets, binned so that there were at least 25 counts in each
bin. Typically, extremely few spectral counts were detected above 2~keV in the
2002 May data, and above 1~keV in the earlier observations, when the
source was fainter.  This binning results in four bins for the 1999 December 
observation, when the source was faintest. 
To verify that the determination of $N_{\rm H}$ was not
affected by calibration uncertainties, we also (a) ran our spectral
analysis excluding data below 0.6~keV (to avoid the Oxygen edge) and
(b) added an edge to the models.  In all cases, we obtained
consistent results.

The RS model is a bad fit for all the data sets. Blackbody, disk-blackbody and
bremsstrahlung all give similar quality fits. Best-fit  kT is in the range of 
$\sim$150-90~eV for the blackbody model, 110-220~eV for the disk-blackbody,
and 120-370~eV for the bremsstrahlung. The power-law model is acceptable for the first and third datasets,
but gives a bad fit for the higher statistics 2002 May data; it is also marginal for the
2001 December data. In general power-law indeces tend to be very large, reflecting the
very soft spectra.   These results, including the best fit
$\chi^2$ and the degrees of freedom (number of bins minus fit parameters)
 are summarized in Table~2, where the errors are 1$\sigma$ for 1
interesting parameter. To verify that different binning did not affect our results in our 
lowest count rate dataset (1999 December), we also analyzed these data with 
15 counts per bin, as in Zezas et al.\ (2002a), obtaining
consistent results. 
  
The best-constrained results are from the 2002 May observation, when
the source was most luminous: the temperature then was
$kT\sim90$~eV for the blackbody model, 110~eV and 120~eV for the disk-blackbody 
and the bremsstrahlung models, respectively. The fits also suggest a large absorption column $N_{\rm H}
\sim 3-5\times 10^{21} \rm cm^{-2}$, depending on the model.  For the rest of this paper 
we will assume that the emission is optically thick, and for simplicity use the blackbody
model. Use of the disk-blackbody model would not change our conclusions.
Figure~3 shows the data, best-fit
blackbody models, and residuals. Figure~4 shows the 1$\sigma$
2-parameter $N_{\rm H}$--$kT$ confidence contours. 
For the 2002 May data we also show the 99\% confidence
contour. Fitting a two-component model (blackbody or disk-blackbody plus
power-law) did not produce a significant improvent of the fit.

The observed best-fit source luminosity (0.1--2~keV) varies from
1.7$\times 10^{38}$\ergs\ in 1999 December to 8$\times
10^{38}$\ergs\ in 2002 May (Table~3). These values can be considered to be lower limits
to the intrinsic source luminosity, since no extinction correction has
been applied. The intrinsic (i.e., emitted) best-fit luminosity is
significantly larger in 2002 May, because of the large $N_{\rm H}$ required
by the fit, and reaches 1.4$\times 10^{40}$\ergs.  In 2002 May, we
estimate a minimum intrinsic luminosity of $\sim 4 \times 10^{39}$\ergs\,
by calculating the flux for the 1$\sigma$ lower limit on both $kT$ and
$N_{\rm H}$. If the emission of the SSS is due to nuclear burning on a white
dwarf (WD) surface (van den Heuvel et al.\ 1992), a hot WD atmosphere
may be more physical than a simple blackbody.  However,
based on the results of Swartz et al.\ (2002), we
estimate that the adoption of a model WD atmosphere will result in
less than a factor of 2 change in the estimated luminosity.

\section{Discussion}

What is this supersoft, luminous, variable source? The large absorption
column and the lack of an obvious identification in the {\it HST\,}
WFPC2 data (Zezas et al.\ 2002b) exclude a foreground object. Given the
low density of supersoft AGNs (Puchnarewicz 1998), the likelihood that this
source could be a background object is low.  If this source belongs to
The Antennae, it is the most luminous galaxian SSS ever detected. 
Our observations suggest that the source has a thermal spectrum. Although
the uncertainties are large, the best-fit temperature does not seem to
increase with luminosity. Taken at face value, this behavior is incompatible
with isotropic emission from a constant radiating area. In the following
analysis, we investigate these trends systematically.

King \& Puchnarewicz (2002) show that blackbody emission from a
region of size $r$ times the Schwarzschild radius of a mass $M$ obeys
the relations
\begin{equation}
L_{\rm sph} = {L\over b} = {2.3\times 10^{44}\over T_{100}^4}{l^2\over
pbr^2}\ {\rm erg\ s^{-1}}, 
\label{l}
\end{equation}
\begin{equation}
M = {1.8\times 10^6\over T_{100}^4}{l\over pr^2} {\rm M}_{\odot},
\label{m}
\end{equation}
where $L_{\rm sph}$ is the inferred isotropic luminosity of the blackbody,
$L$ the true
source luminosity,  $b$ the beaming factor that accounts for eventual
non-spherical emission, $l = L/L_{\rm Edd}$ the Eddington factor, $p \sim 1$
a measure of the geometrical deviation from a spherical photosphere, and
$T_{100}$ the source temperature $T$ in units of 100 eV.

Dividing (\ref{l}) by (\ref{m}) gives $lm/b$ in terms of $L_{\rm sph}$,
where $m = M/M_{\odot}$.  Dividing (\ref{l}) by the square of eq.\ (2)
gives $pr^2m^2/b$ in terms of $L_{\rm sph}T^{-4}$, and we can deduce
$l/mpr^2$ in terms of $T^4$. We calculate these quantities (listed in
Table~3) from the results of Table~2 for the blackbody spectrum.   
It is easy to show from Table~1 that $(lm/b) \propto (pr^2m^2/b)^{2/3}$,
so that
\begin{equation}
r \sim l^{3/4}p^{-1/2}b^{-1/4}.
\label{r}
\end{equation} 
Thus at most one of the three quantities $b, l, pr^2$ can be constant. We
consider three cases.

{\it (i) Near-isotropic emission from an intermediate-mass black hole}
($b \sim 1 \sim p$). Given the peak luminosity of the SSS, this case
requires a black-hole mass $\geq$100\msun (cf Miller et al.\ 2003)
However, the large increase (a factor $\sim$1000) in radiating area is
very hard to reconcile with a simple picture in which a black hole
accretes from an accretion disk and the blackbody emission comes from
a region of a few Schwarzschild radii. Column 3 of Table~3 shows that
$l$ must increase by a factor 81 between the 1st and 4th
observations; for this type of geometry we clearly have $p \sim 1$, so
the radius factor $r$ must increase by a factor $\sim$25.  Since the
blackbody emission comprises most of the putative accretion
luminosity, it must come from deep within the potential well, i.e., $r
\la$ a few. So, if we (generously) set $r \sim 3$ in Observation 4,
the emission must be confined to an implausibly small region of only
0.1 Schwarzschild radii in Observation 1.

Although the backbody fits point to the exclusion of an massive 
black hole, this conclusion is not iron-clad. It hinges 
on the assumption that the 1999 December
emission is indeed a blackbody. As discussed earlier, this is the least 
well constrained dataset, so that we cannot exclude that the spectrum may
follow a power-law distribution. Although the  power-law $\Gamma \sim 4$ we obtain 
for the 1999 December data is extremely steep, in excess of typical black-hole binary power-laws,
the uncertainties are large (see also Zezas et al 2002a).
Power-law components have been seen to dominate the emission in ULXs in
low state (see, e.g. La~Parola et al.\ 2001; Kubota et al.\ 2001; Fabbiano et al 2003).
If the first observation is discarded from the blackbody analysis
of Table 2, the required radius increase between observations 2 and 4
is reduced to a factor $\sim 2$. If we only consider observations 3 and 4,
we obtain a radius increase of a factor of $\sim 4$. It remains to be demonstrated how
this smaller increase would fit into an intermediate--mass
black hole picture.

{\it (ii) Varying beaming at constant luminosity}. 
An opposite extreme from the near-isotropic case is
$l \sim$ constant, in which case $b$ decreases by a factor
$\sim$139 between Observations 1 and 4. The last column of Table~1
then shows that $p$ must simultaneously increase by a factor $\sim$10.
Physically, holding $l$ constant while other quantities vary widely is
plausible only in one situation, namely that the source is radiating
constantly at the Eddington limit ($l = 1$) while the accretion rate may
change. In this case the range of $b$ is $m - m/185$. The $l/mpr^2$
column of Table~1 then shows that $pr^2 = 2.7\times 10^6m^{-1}$ in
Observation 4, giving the radiating object a radius of
$R = 4.8\times 10^8m^{1/2}p^{-1/2}~{\rm cm}$.
For $m \sim 1$, $p \sim 1$ this radius is suggestively close to the radius
of a white dwarf. This is not surprising, as the inferred
temperature and luminosity are now typical of supersoft X--ray
binaries, which are thought to be powered by nuclear burning of matter
accreting on to a white dwarf.  

If the SSS is an accreting WD, our results would be consistent with an
increasingly super-Eddington accretion flow ($\dot M \ga \dot M_{\rm
Edd} \sim 10^{-7}$\msun {\rm yr}$^{-1}$ for steady nuclear burning on
a WD surface). As $\dot M$ increases beyond $\dot M_{\rm Edd}$, the
flow geometry apparently changes gradually from a thin disk (1999
December), with burning on a narrow, isotropically emitting ($b \sim
1$) equatorial band ($p \sim 0.1$) on the white dwarf, to a thick disk
with burning on most of the WD surface ($p \sim 1$), but now with a
strongly anisotropic radiation pattern ($b \sim 10^{-2}$) in our
direction. The most likely cause of this anisotropy is warping of
the accretion disk. We note that disks are known to warp in supersoft
X-ray binaries (Southwell et al.\ 1997), essentially because of their
proximity to the Eddington limit.  At still higher $\dot M$ nuclear
burning on white dwarfs drives either envelope expansion where the
source swells up to red-giant dimensions, or a vigorous wind outflow,
or both (cf Hachisu et al., 1996). In all cases the system is likely
to be extinguished as an X-ray source.

{\it (iii) Mildly anisotropic emission ($ b\ga 0.1$) from a
stellar-mass black hole.} At first sight the photospheric radii of
$\sim$10$^8$--10$^9$~cm we deduce from our observations do not appear
any more natural for a stellar-mass black hole than for an
intermediate-mass one as in (i) above. However, Mukai et al.\ (2002)
have pointed out that accretion at rates comparable to Eddington must
lead to outflow, and have shown that the electron scattering opacity of
the resulting wind does imply supersoft emission with a photospheric
size of this order. The M101 source studied by Mukai et al.\ (2002)
has a supersoft luminosity of order $10^{39}$\ergs\ and, hence, does
not require anisotropic emission for a black hole mass
$\ga$10\msun. However, their analysis is easily extended to the case
that an Eddington-limited source blows out a wind confined to a double
cone of total solid angle $4\pi b$ about the black-hole axis. Since
this wind represents the path of lowest optical depth through the
accretion flow, the radiation will escape this way also, implying $p
\sim b$. We follow Mukai et al.\ (2002) in assuming a constant
velocity for the outflowing material, since this material is likely to
achieve escape velocity and coast thereafter. We neglect any
emission from this wind and  compute the Thomson optical depth
$\tau$ by integrating the electron density $N_e$ from radius $R$ to
infinity. Since $\int N_e dr \simeq \dot M_{\rm out}/4\pi bvRm_{\rm
H}$ we find a photospheric radius
\begin{equation} 
R_{\rm ph} = {3\times 10^8\over bv_9}\dot M_{19}\ {\rm cm},
\label{wind}
\end{equation} 
where $v_9$ is $v$ in units of $10^9$ cm~s$^{-1}$ and $\dot M_{19}$ is
the outflow rate in units of $10^{19}$~g~s$^{-1}$, the Eddington
accretion rate for a 10\msun\ black hole. Clearly, we must choose a
mass of this order for consistency. The luminosity
$\sim$10$^{38}$\ergs\ of Observation 1 is then definitely
sub-Eddington, so presumably there is very little outflow and we see
down to the inner accretion disk directly in this
observation. Observations 2, 3, and 4 are all close to $L_{\rm Edd}$,
i.e., $l = 1$. Since $b = p$ and the true blackbody radius $R
= 3\times 10^5 rm$~cm, we can read off the size of the photosphere
directly from the last column of the Table, and $l/b$ also. This gives
an inner disk radius of order $2\times 10^8$~cm for Observation 1,
with $R_{\rm ph} = 2\times 10^9$~cm, $5\times 10^8$~cm, and $5\times
10^9$~cm for Observations 2, 3, and 4, respectively.  Simultaneously
$l/b$ increases from $\sim$0.1 to 3.1, 0.6, and 11. This is consistent
with $l = 0.1$, 1, 1, and 1 over the four observations and $b$
eventually decreasing to a value of $\sim$0.10, presumably as $\dot
M_{\rm acc}$ rises above the Eddington rate.  Self-consistently, the
assumed $v_9$ is above the escape value for the values of $R_{\rm
ph}$.  Of course, $b$ would be larger still if we took a BH mass of
15\msun\ rather than 10\msun, as observed in GRO~J1915+105 (Greiner et
al.\ 2001).

The above values are consistent with the suggestion by King et al.\ (2001)
and King (2002) that ULXs are actually X-ray binaries involving
stellar-mass black holes, but with mildly anisotropic radiation
patterns ($b \sim 0.1$) resulting from
accretion at close to the Eddington rate onto the black hole. 
The above scenario can arise in two cases (i) thermal-time-scale mass transfer when the
companion star in a high--mass X--ray binary fills its Roche lobe, and
(ii) bright outbursts of soft X--ray transients (King, 2002). The
first case is dominant in most galaxies, although the second must
account for ULXs in ellipticals.
Confirmation that this is a
reasonable explanation for ULXs as a class comes from Grimm, Gilfanov,
\& Sunyaev (2002), who show that---when normalized by the star formation
rate---ULXs form a natural extension to the luminosity function of
high-mass X-ray binaries in nearby galaxies.

\section{Conclusions}
We have discovered a variable SSS ($kT\sim 90$~keV) in The Antennae, Source
CXOANT J120151.6$-$185231.9, which reached a peak intrinsic luminosity of
$1.4 \times 10^{40}$\ergs\ in 2002 May.

Near-isotropic emission from an intermediate-mass
black hole accreting from a disk would be incompatible with our observations
of this source in the most likely case of soft thermal emission, 
as the radiating area would have to increase by
more than a factor 1000 over the four observations.
There remains a less likely possibility that the 1999 December
emission may be due to a low-intensity power-law dominated state (e.g. Kubota et al.\
2001), in which case 
the required area increase is reduced to a less demanding factor 10.

A possible solution is a white dwarf with $M \sim 1$\msun, accreting at the
Eddington limit ($l = 1$) and with a variable beaming factor.  
 This explanation has the advantage of giving a natural
scale for the deduced photospheric radius, but does require extreme
beaming (up to $b \sim 10^{-2}$) . 

A second possible solution involves outflow from a stellar-mass black
hole, accreting near the Eddington limit (Mukai et al.\ 2002). A consistent
explanation of our observations results if this hypothesis is combined
with the suggestion
by King et al.\ (2001) that ULXs are actually X-ray binaries involving
stellar-mass black holes, but with mildly anisotropic radiation
patterns ($b \sim 0.1$).  

The stellar-mass black hole solution is the more conservative choice
in the present case, as it does not require as extreme an anisotropy as
a white dwarf. However the latter may remain a realistic candidate for
slightly less luminous supersoft ULXs.

\acknowledgments

 We thank the CXC DS and SDS teams for their efforts in reducing the
data and developing the software used for the reduction (SDP) and
analysis (CIAO). We thank John Raymonds for comments on this paper and
Larry David for advice on ACIS spectral
analysis at low energies. We are grateful to Aya Kubota for her careful reading 
of this paper.  This work was supported by NASA contract
NAS~8--39073 (CXC) and NASA Grant G02-3135X.  ARK gratefully
acknowledges a Royal Society Wolfson Research Merit Award.
We are grateful to the Aspen Center for Physics for fostering the
collaboration that resulted in this work.

{}

\makeatletter
\def\jnl@aj{AJ}
\ifx\revtex@jnl\jnl@aj\let\tablebreak=\nl\fi
\makeatother
\begin{deluxetable}{ccc}
\tabletypesize{\scriptsize}
\tablecolumns{3}
\tablewidth{0pt}
\tablecaption{Observations }
\tablehead{ \colhead{OBSID} & \colhead{Date} & \colhead{Net Exp. }
\\
\colhead{} & \colhead{} & \colhead{(Ks)} }
\startdata
315 & 1999-12-01 & 75 \\
3040 & 2001-12-29 & 64  \\
3043 & 2002-04-18 &  61 \\
3042 & 2002-05-31 & 67\\
\enddata
\end{deluxetable}

\makeatletter
\def\jnl@aj{AJ}
\ifx\revtex@jnl\jnl@aj\let\tablebreak=\nl\fi
\makeatother
\ptlandscape
\begin{deluxetable}{lcccccccccccccccccc}
\tabletypesize{\scriptsize}
\rotate
\tablecolumns{17}
\tablewidth{0pt}
\tablecaption{ Spectral Fits}
\tablehead{ 
  \colhead{OBSID} &       \multicolumn{3}{c}{B. Body}  &  \multicolumn{3}{c}{Disk-BB} & \multicolumn{3}{c}{Bremss.}  & \multicolumn{3}{c}{RS}  & \multicolumn{3}{c}{PO} \\               
 \colhead{}    & \colhead{kT}  & \colhead{\nh}  & \colhead{\x2}  &  \colhead{kT$_{in}$}  & \colhead{\nh}  & \colhead{\x2}  & \colhead{kT}  & \colhead{\nh}  & \colhead{\x2}  &  \colhead{kT}  & \colhead{\nh}  & \colhead{\x2}  &\colhead{$\Gamma$}  & \colhead{\nh}  & \colhead{\x2}  \\
  \colhead{}       & \colhead{keV} & \colhead{$10^{22}~cm^{-2}$} & \colhead{d.o.f.}&
\colhead{keV} & \colhead{$10^{22}~cm^{-2}$} &  \colhead{d.o.f.} & \colhead{keV} &
\colhead{$10^{22}~cm^{-2}$} & \colhead{d.o.f.}& \colhead{keV} & \colhead{$10^{22}~cm^{-2}$} & \colhead{d.o.f.}
& &\colhead{$10^{22}~cm^{-2}$} & \colhead{d.o.f.} }
\startdata
315  & $0.15^{+0.03}_{-0.04}$  &  $<0.12$ & 1.3 &
	$0.22^{+0.04}_{-0.09}$  &  $<0.15$ & 0.43 &
       $0.37^{+0.27}_{-0.19}$  &  $0.04 (<0.2)$  & 0.3 &
        $0.22$                  &  $<0.04$        & 13.4 &  
         $4.74^{+5.3}_{-1.3}$    &  $0.18^{+0.57}_{-0.13}$ & 0.23\\
        & & & 1&   & & 1& & & 1& & & 1& & & 1 \\
3040    & $0.10^{+0.02}_{-0.024}$  &  $0.20^{+0.22}_{-0.18}$ & 4.0 &
        $0.11^{+0.04}_{-0.03}$  &  $0.25^{+0.22}_{-0.15}$ & 3.8 &
           $0.14^{+0.05}_{-0.05}$  &  $0.32^{+0.32}_{-0.15}$ & 3.6 &
          $0.09^{+0.05}_{-0.03}$  &  $0.63^{+0.19}_{-0.10}$ & 11.4 &
         $9.6 (>7.8) $           &  $0.72^{+0.6}_{-0.2}$   & 6.18 
           \\
           & & & 4&   & & 4& & & 4& & & 4& & & 4\\
3043   & $0.13^{+0.02}_{-0.04}$  &  $0.025 (<0.22)$        & 2.2 &
           $0.14^{+0.05}_{-0.04}$  &  $0.09 (<0.29)$         & 1.91 &
           $0.20^{+0.10}_{-0.08}$  &  $0.15^{+0.05}_{-0.11}$ & 1.64&
           $0.20$                  & $<$0.1                    & 18.5 &
           $2.64^{+2.0}_{-2.0}$    &  $0.46^{+0.08}_{-0.21}$ & 2.17 \\
           & & & 3&   & & 3& & & 3& & & 3& & & 3\\
3042 &  $0.09^{+0.02}_{-0.01}$  & $0.28^{+0.22}_{-0.16}$  & 21.7&  
           $0.11^{+0.02}_{-0.03}$  & $0.35^{+0.23}_{-0.15}$  & 22.4&
           $0.12^{+0.05}_{-0.03}$  & $0.49^{+0.28}_{-0.21}$  & 23.5 &
           $0.26$                  & $<0.06$                 &44.7&
           $9.0 (>7.1)$            & $0.70^{+0.1}_{-0.19}$   &39.2\\
           & & & 10&   & & 10& & & 10& & & 10& & & 10\\
\enddata
\end{deluxetable}

\makeatletter
\def\jnl@aj{AJ}
\ifx\revtex@jnl\jnl@aj\let\tablebreak=\nl\fi
\makeatother
\begin{deluxetable}{cccccc}
\tabletypesize{\scriptsize}
\tablecolumns{6}
\tablewidth{0pt}
\tablecaption{Blackbody Luminosities and Derived Parameters}
\tablehead{ \colhead{OBSID}
& 
\colhead{$\rm{L_{X, 0.1-2~\rm keV}^{observed}}$} 
& \colhead{$\rm{L_{X, 0.1-2~\rm keV}^{emitted}}$}& \colhead{$lm/b$} & \colhead{$l/mpr^2$} &
\colhead{ $pr^2m^2/b$}\\
\colhead{}  &
\colhead{$10^{39}$\ergs} & \colhead{$10^{39}$\ergs}
& \colhead{} & \colhead{} & \colhead{}\\
\colhead{(1)} & \colhead{(2)} & \colhead{(3)} &\colhead{(4)} & \colhead{(5)} & \colhead{(6)}
 }
\startdata
315 &   0.17 & 0.17& 1.33           &  $2.8\times 10^{-6}$ & $4.8\times 10^5$ \\
3040 &   0.43 & 3.9& 30.5           &  $5.6\times 10^{-7}$ & $5.4\times 10^7$\\
3043 &   0.53 & 0.8& 6.2           &  $1.6\times 10^{-6}$ & $3.9\times 10^6$\\
3042 &   0.80 & 13.8 & 108            &  $3.6\times 10^{-7}$ & $2.9\times 10^8$\\
\enddata
\end{deluxetable}


\setcounter{figure}{0}

\begin{figure}
\includegraphics[width=12.0cm]{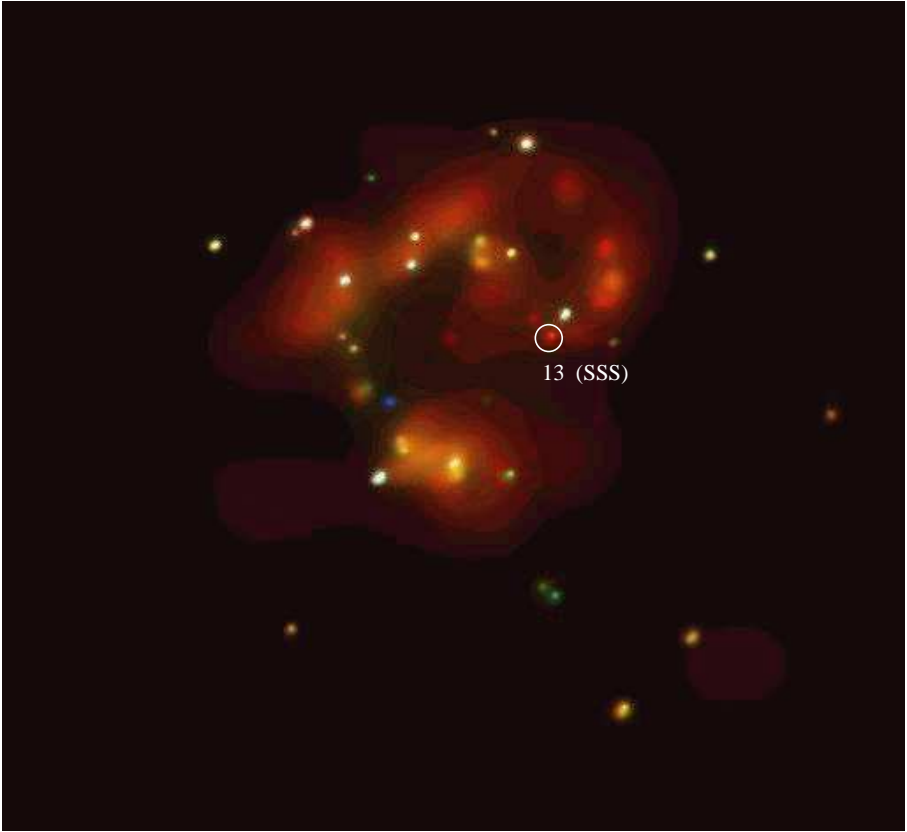}
\caption{The Dec 1999 image of The Antennae (Fabbiano et al.\ 2001), with 
the super soft ULXs discussed in this paper identified by a circle and source
number from Zezas et al.\ (2002a). Soft emission is red, hard emission is blue.
Luminous sources appear white because they emit in the entire spectral band.}
\end{figure}

\begin{figure}
\begin{tabular}{cc}
\rotatebox{270}{\includegraphics[width=7.0cm]{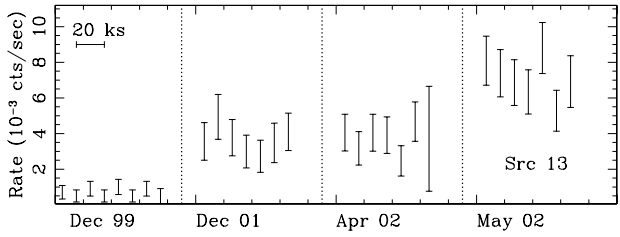}}
\end{tabular}
\caption{The light curve of CXOANT J120151.6$-$185231.9. The starting date
of each observation is given in Table 1; the data are binned in 10~ks time
bins.}
\end{figure}

\begin{figure}
\begin{tabular}{cc}
\rotatebox{270}{\includegraphics[width=7.0cm]{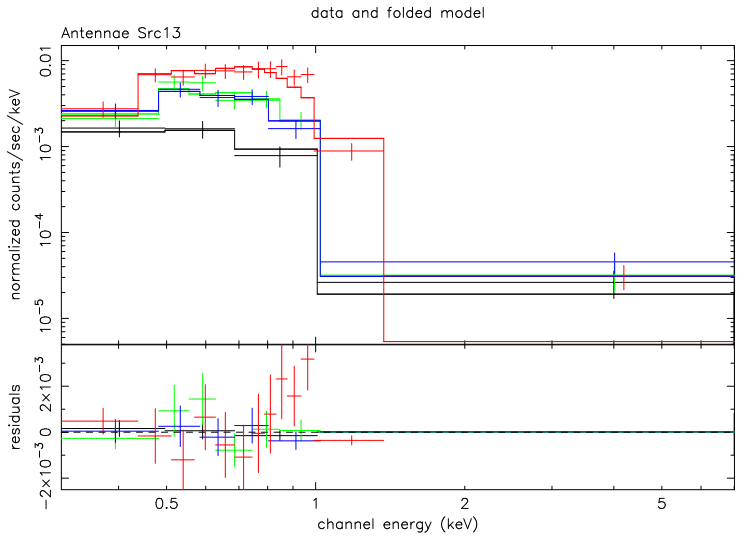}} & \\
\end{tabular}
\caption{Observed spectral data (see text) and best-fit blackbody spectra, with
the fit residuals. Black: 1999 December; green: 2001 December; blue: 2002
April; red: 2002 May. We show only the 0.3 -- 2~keV range, where the greatest 
majority of counts were detected. In all cases the last bin used for the fitting extends to 7~keV.}
\end{figure}

\begin{figure}
\begin{tabular}{cc}
\rotatebox{270}{\includegraphics[width=7.0cm]{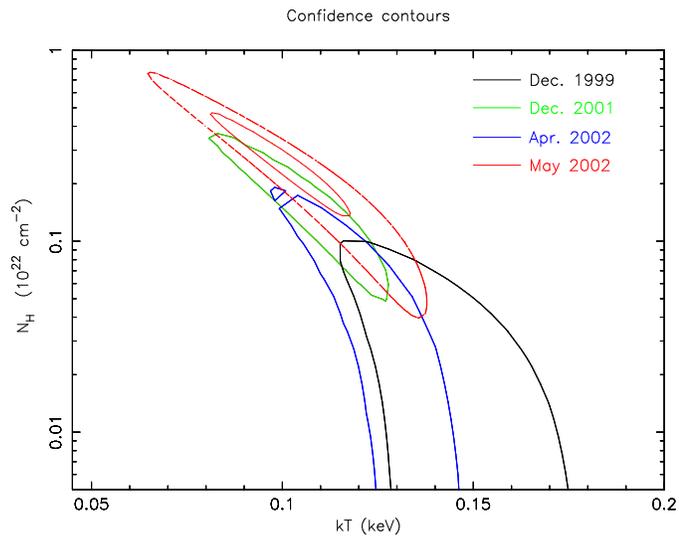}} & \\
\end{tabular}
\caption{$N_{\rm H}$--$kT$ confidence contours. Solid contours are at 1$\sigma$ for
two interesting parameters. For the 2002 May observations, the dashed contour
is at 99\% for two interesting parameters.}
\end{figure}

\end{document}